\documentclass[preprint,12pt]{elsarticle}

\usepackage{amssymb}

\def\class#1{{\tt #1}}
\def\e{{\rm e}}
\def\d{{\rm d}}
\def\un#1{{\,\rm #1}}

\journal{Computer Physics Communications}

\begin{document}

\begin{frontmatter}

\title{Elegent -- an elastic event generator}

\author{J.~Ka\v spar\corref{cor1}\fnref{fn1}}
\ead{jan.kaspar@cern.ch}
\address{CERN, 1211 Geneva 23, Switzerland}

\cortext[cor1]{Corresponding author}
\fntext[fn1]{On leave of absence from Institute of Physics, AS CR, v.v.i., 182 21 Prague 8, Czech Republic.}

\begin{abstract}
Although elastic scattering of nucleons may look like a simple process, it presents a long-lasting challenge for theory. Due to missing hard energy scale, the perturbative QCD can not be applied. Instead, many phenomenological/theoretical models have emerged. In this paper we present a unified implementation of some of the most prominent models in a C++ library, moreover extended to account for effects of the electromagnetic interaction. The library is complemented with a number of utilities. For instance, programs to sample many distributions of interest in four-momentum transfer squared, $t$, impact parameter, $b$, and collision energy $\sqrt s$. These distributions at ISR, S$\rm p\bar p$S, RHIC, Tevatron and LHC energies are available for download from the project web site. Both in the form of ROOT files and PDF figures providing comparisons among the models. The package includes also a tool for Monte-Carlo generation of elastic scattering events, which can easily be embedded in any other program framework.
\end{abstract}


\end{frontmatter}

\newpage


\section{Introduction}\label{s:inc}

The elastic scattering of nucleons (i.e.~protons or anti-protons) may look like their simplest interaction, however, it still presents a challenge for theory to describe it. The complication may be seen in the fact that the coupling constant of quantum chromodynamics (QCD) becomes large at low energy scales (low momentum transfer is characteristic for the elastic scattering of nucleons). Consequently, one can not apply the straight-forward perturbative calculations like in quantum electrodynamics (QED) for example. Instead of describing the elastic scattering from first principles, many model descriptions have been developed. These more or less QCD-motivated models are often built on Regge theory and/or eikonal formalism grounds. As there are new experimental efforts at the LHC \cite{totem,alfa}, we have felt it useful to collect some of the most prominent models in an easily accessible library that could be used for various applications (C++ code, comparison to data, Monte-Carlo generator, etc). This became the core of the Elegent (ELastic Event GENeraTor) package.

Out of the three fundamental forces relevant for particle interactions only the strong and electromagnetic forces are important for elastic scattering of nucleons. The four-momentum transfer squared, $t$, typical for the elastic scattering is $|t| \lesssim 10\un{GeV^2}$. Compared to that, the carriers of the weak forces are much heavier, thus the influence of the weak interaction is negligible. Some models of the strong interaction (traditionally called {\em hadronic models}) are described in Section \ref{s:had mod}, the electromagnetic (traditionally called {\em Coulomb}) interaction is discussed in Section \ref{s:coul mod}. Some approaches to evaluate the interference effects between the hadronic and Coulomb forces are reviewed in Section \ref{s:int mod}.

Although nucleons are spin $1/2$ particles, there are indications that for forward elastic scattering only one spin amplitude is dominant (for details see \cite[section 1]{jan_thesis}). This sets the traditional theoretical framework where only one scattering amplitude is considered.


\section{Hadronic models}\label{s:had mod}

The following models are currently implemented in the Elegent package, for a more detailed model review see \cite[section 1.1]{jan_thesis} or \cite[section 4]{dremin13}.

The {\bf model of Block et al.}~\cite{bh99,block06} is a QCD-inspired model formulated in eikonal formalism . The eikonal receives four contributions due to quark-quark, quark-gluon, gluon-gluon interactions and an odderon exchange. The latter is responsible for the difference between $\rm pp$ and $\rm \bar p p$ scattering and is negligible at LHC energies. Each of the eikonal contributions is factorised into energy dependence term (proportional to the integral cross-section of a given sub-process) and impact-parameter profile. All four profiles have the same form but are scaled with a parameter reflecting the areas occupied by quarks and gluons in the nucleon. The gluon-gluon cross-section is modelled after QCD, adopting a gluon distribution function that behaves as $1/x^{1+\epsilon}$ at low momentum fractions $x$. It is this high soft-gluon content that is responsible for the cross-section rise with energy. This model is implemented as class \class{BHModel} in the library source code, using parameter values from publication \cite{block06}.

The {\bf model of Bourrely et al.}~\cite{bsw79,bsw84,bsw03,bsw11} is also formulated within eikonal description. The eikonal includes two terms corresponding to two scattering mechanisms: pomeron and reggeon exchange. The pomeron term is factorised into a product of two functions, one depending on centre-of-mass energy, $\sqrt s$, and one on impact parameter, $b$. The energy dependence is deduced from asymptotic quantum field theory behaviour \cite{wu70}. The impact-parameter profile of the pomeron exchange is derived by assuming similar distributions of electric charge and nuclear matter in the nucleon. The charge distribution is extracted from electromagnetic form factor and is further modified by a slowly varying function in order to get the complete profile function. The reggeons considered in this model include $\rm A_2$, $\rho$ and $\omega$. The trajectories are described with a traditional parametrisation with $s^{\alpha - 1}$ energy dependence and $\e^{-bt}$ momentum-transfer dependence. The reggeon contribution is responsible for the $\rm pp$ and $\rm \bar pp$ difference at low energies and is negligible at LHC and higher energies. This model is implemented as class \class{BSWModel} in the library source code, using parameter values from publication \cite{bsw03}.

In the {\bf model of Islam et al.}~the proton is pictured in an effective quantum field theory model (gauged linear sigma model) \cite{islam06}. The proposed soliton solution has a mass comparable to the proton and divides the proton to two distinct layers: an outer cloud of quark-antiquark condensate and an inner core of topological charge. Later on \cite{islam05}, a third, the inner-most layer has been added: a valence-quark bag. These three layers give rise to three mechanisms contributing to elastic proton-proton scattering, each being dominant at a different region of $t$. At the lowest $|t|$ it is the diffraction scattering originating from a glancing overlapping of the outer clouds \cite{islam84,islam87}, at medium $|t|$ the core-core scattering \cite{islam06} and at high $|t|$ the valence-quark scattering \cite{islam05,islam09}. The diffraction scattering is described in impact-parameter picture using a Fermi profile function. As a consequence of the underlying quantum field theory model, the core-core scattering is mediated by the $\omega$ meson behaving as an elementary particle. The corresponding amplitude thus contains a Feynman propagator and a form factor ensuring an exponential fall off at higher $|t|$ values. Regarding the valence-quark scattering, the model comes with two alternative mechanisms: interaction via a BFKL gluon ladder (denoted {\em hard pomeron} variant, HP) \cite{islam05} or interaction mediated by low-$x$ gluons surrounding the valence quark (variant {\em low-x gluons}, LxG) \cite{islam09}. This model is implemented as class \class{IslamModel} in the library source code, using parameter values from publication \cite{islam06} and \cite{islam09} for the LxG extension.

In the {\bf model of Jenkovszky et al.}~\cite{jenkovszky11} the elastic scattering of nucleons is attributed to $t$-channel exchanges of a pomeron, an odderon and reggeons $\rm f$ and $\omega$. The reggeon-exchange amplitudes have the traditional form with $s^{\alpha - 1}$ energy dependence and residua $\e^{-bt}$ and are negligible at LHC and higher energies. The pomeron and odderon amplitudes have a more complicated form arising from the assumed double Regge pole nature of the pomeron \cite[section 2]{jenkovszky86}. Moreover, the model comes with several parametrisations of the pomeron Regge trajectory $\alpha(t)$. The linear parametrisation is discussed in the greatest detail in the publication and is the only implemented in Elegent at the moment. This model is implemented as class \class{JenkovszkyModel}, using parameter values from publication \cite{jenkovszky11}.

The {\bf model of Petrov et al.}~\cite{petrov02} describes the collision of nucleons as an exchange of reggeons: pomerons, an odderon, $\rm f$ and $\omega$. Each of these trajectories contributes to the eikonal with a term proportional to $s^{\alpha - 1}$ and with a Gaussian impact-parameter profile $\e^{-c b^2}$. The model has two variants: including two pomerons (denoted {\em 2P}\/) and three pomerons ({\em 3P}\/). This model is implemented as class \class{PPPModel}.

A {\bf simple exponential model} with a constant phase is implemented (class \class{ExpModel}) for convenience and comparison.


\section{Coulomb amplitudes}\label{s:coul mod}

In Elegent, the Coulomb scattering amplitude is calculated in the framework of QED, within the leading approximation (one photon exchange, OPE). As argued in \cite[section 1.3.6]{jan_thesis}, there are reasons to believe that this approximation is sufficient in the domain of very low momentum transfers where the Coulomb amplitude dominates the hadronic one. One should also note that higher order corrections (multi photon exchange) go beyond the scope of pure QED as excited states of the nucleon need to be considered in loop diagrams too (for more details see \cite{arrington07,puckett10}).

The QED framework can naturally account for the nucleon structure and its anomalous magnetic moment by introducing electric and magnetic form factors that modify the standard QED vertex function. Moreover, at high energies (like at LHC) and low momentum transfers ($|t| \lesssim 10\un{GeV^2}$), the electric and magnetic form factors can be merged into one effective form factor $G_{\rm eff}(t)$ \cite[equation (31)]{block06}. The differential cross-section than reads:
\begin{equation}
{\d\sigma\over\d t} = {4\pi \alpha^2\over t^2}\, G_{\rm eff}^4(t)\ ,
\end{equation}
where $\alpha$ is the fine-structure constant.

In the Elegent package, the user can choose (via {\tt CoulombInterference::\discretionary{}{}{}FFType}) between several form factor parametrisations. They account for increasing amount of experimental data as well as progressing theoretical development. One of the first publications is by Hofstadter \cite{hofstadter58} who used a dipole parametrisations. This was later updated by Borkowski et al.~\cite{borkowski74,borkowski75} parametrising both form factors by a sum of four poles. Further publications Kelly \cite{kelly04}, Arrington et al.~\cite{arrington07} and Puckett \cite{puckett10} use a parametrisation satisfying dimensional scaling at high $|t|$. A graphical comparison of these form factors is available in \cite[Figure~1.7]{jan_thesis}.

The Coulomb amplitude and nuclear form factors are implemented in class {\tt CoulombInterference}.


\section{Coulomb-hadronic interference}\label{s:int mod}

After introducing hadronic and Coulomb amplitudes, one needs to consider the effects of their interference in order to complete the picture. Historically, there are two ways to calculate these effects: in Feynman-diagram framework and using eikonal approximation.

The Feynman-diagram approach is best represented by the work of {\bf West and Yennie} \cite{wy68} (a summary is available in \cite[section 1.3.4]{jan_thesis}, also mentioning the preceding works). The basic idea is to evaluate diagrams including both Coulomb and hadronic interactions, treating correctly the infrared divergences due to the zero photon mass. Moreover, these diagrams may contain loops, the evaluation of which requires the off-shell hadronic amplitudes -- that are generally (and particularly for the models described in Section \ref{s:had mod}) not known. Therefore, the authors made a number of simplifying approximations (e.g.~slow variation of the hadronic amplitude phase) in order to derive an interference formula in a closed form. This formula is denoted {\em WY} and -- strictly speaking -- is internally inconsistent for the models from Section \ref{s:had mod} with non-constant phase. The authors made one more simplifying step -- assuming purely exponential decrease of the hadronic amplitude with $|t|$ -- and obtained a simple interference formula, denoted {\em SWY}. Although this simplification is inconsistent with many models, this formula has been often used until recently and therefore it is still implemented in the Elegent package.

The eikonal calculation was pursued by a number of authors (reviewed in \cite[section 1.3.5]{jan_thesis}), most recently by {\bf Kundr\' at and Lokaj\' i\v cek} \cite{kl94}. This approach is based on the additivity of eikonals: the complete eikonal is given by the sum of Coulomb eikonal (obtained from the OPE Coulomb amplitude) and the hadronic eikonal (given by Fourier-Bessel transform of the hadronic amplitude). Within the leading approximation in $\alpha$, the authors derived a closed-form interference formula (denoted {\em KL}) without explicitly assuming anything about the hadronic amplitude.

In both approaches above, the interference formulae are derived in the leading $\alpha$ order only. In case of the eikonal approach, however, it is possible to perform a numerical calculation to all orders. As shown in \cite[section 1.3.7]{jan_thesis}, the KL formula is very accurate for small $|t|$ up to the (first) dip, where the disagreement peaks at the level of $1\un{\%}$.

Both approaches consider only certain (classes of) diagrams contributing to the Coulomb-hadronic interference. The effects of the neglected diagrams -- as discussed in \cite[section 1.3.6]{jan_thesis} -- can become significant for $|t|\gtrsim 1\un{GeV^2}$. Therefore, one should use the interference formulae with care.

The Coulomb-nuclear interference models are implemented in class\break {\tt CoulombInterference}, the choice of the formula is made via its {\tt mode} variable.


\section{Program package}\label{s:prog}

The core of the Elegent package is a C++ library implementing the physics models from Sections \ref{s:had mod}, \ref{s:coul mod} and \ref{s:int mod}. This core library is described in Section \ref{s:prog lib}. Furthermore, the package contains a number of convenience utilities, see Section \ref{s:prog util}.

Further practical-oriented information, implementation details and users' manual can be found at the project web page \cite{elegent}.

\subsection{Model library}\label{s:prog lib}

In the library source code, the four-momentum transfer squared $t$ is treated as negative number in $\rm GeV^2$. The units of the impact parameter $b$ are $\rm fm$.

The class {\tt Math} collects numerical integration routines.

The class {\tt Constants} encapsulates physics and mathematical constants (e.g.~proton mass, Euler's constants) as well as process description (e.g.~$\rm pp$ or $\rm \bar pp$ mode, centre-of-mass energy). The constants are set up by the static {\tt Init} method, creating an instance of the class referenced by the global pointer {\tt cnts}.

The class {\tt Model} is the mother class for all the hadronic models described in Section \ref{s:had mod}. Each model object has two labels: {\tt fullLabel} (intended for figure legends) and {\tt shortLabel} (intended for object names in ROOT files). Each of the labels have this structure: {\tt name:mode (variant) [version]}. The {\tt mode} (e.g.~which amplitude components are included) and {\tt variant} are optional entries and {\tt version} gives a reference to the paper(s) used for the implementation. The mode and variant can be chosen by the {\tt Configure} method. Afterwards, the {\tt Init} method shall be called to set up appropriate values of model parameters and to perform (potentially) time expensive operations (e.g.~pre-sampling of certain distributions). The state of a model can be queried by calling the {\tt Print} method. The {\tt Amp(t)} method calculates the scattering amplitude $F(t)$ using the following normalisation:
\begin{equation}
{\d\sigma\over\d t} = {\pi\over s p^2} |F(t)|^2\ ,\qquad \sigma_{\rm tot} = {4\pi\over p\sqrt s} \Im F(t = 0)\ ,
\end{equation}
where $s$ denotes the square of the collision energy and $p$ is the momentum of a colliding nucleon in the centre-of-mass frame. The profile function $A(b)$ (i.e.~impact-parameter amplitude) is calculated by the {\tt Prf(b)} method. The profile function is normalised according to these relations:
\begin{equation}
F(t) = 2 p \sqrt s \int b\, \d b\, J_0(b \sqrt{-t})\, A(b)\ ,\quad A(b) = {1\over 4p\sqrt s} \int \d t\, J_0(b\sqrt{-t})\,F(t)\ ,
\end{equation}
where $J_0$ stands for the Bessel function. There is a global model pointer {\tt model} referring to the actual/working model.

The class {\tt CoulombInterference} implements the models from Sections \ref{s:coul mod} and \ref{s:int mod}. The form factor parametrisation is set via the {\tt ffType} option. The methods {\tt FF\_e(t)}, {\tt FF\_m(t)} and {\tt FF\_sq(t)} return the electric, magnetic and square of the effective form factor. Scattering amplitudes can be calculated by {\tt Amp(t)} method. The type of the amplitude is chosen via the {\tt mode} variable: {\tt mPC} (pure Coulomb amplitude), {\tt mPH} (pure hadronic amplitude), {\tt mWY} (West-Yennie formula), {\tt mSWY} (simplified West-Yennie formula) and {\tt mKL} (Kundr\'at-Lokaj\' i\v cek formula). The hadronic model used for Coulomb interference calculations is given by the global variable {\tt model} introduced above. For convenience, the class implements methods to calculate several derived quantities of interest. The {\tt R(t)} method evaluates relative difference between the SWY and KL formulae
\begin{equation}\label{e:R}
R(t) = {|F^{\rm KL}(t)|^2 - |F^{\rm WY}(t)|^2 \over |F^{\rm KL}(t)|^2}\ ,
\end{equation}
the {\tt Z(t)} method gives the importance of the interference term
\begin{equation}\label{e:Z}
Z(t) = {|F^{\rm KL}(t)|^2 - |F^{\rm PH}(t)|^2 - |F^{\rm PC}(t)|^2 \over |F^{\rm KL}(t)|^2}
\end{equation}
and the {\tt C(t)} measures the influence of Coulomb interaction to the differential cross-section
\begin{equation}\label{e:C}
C(t) = {|F^{\rm KL}(t)|^2 - |F^{\rm PH}(t)|^2 \over |F^{\rm PH}(t)|^2} \ .
\end{equation}
There is a global pointer {\tt coulomb} to an instance of the {\tt CoulombInterference} class.

\subsection{Utilities}\label{s:prog util}

\begin{table}
\def\Box#1{\leftskip0pt\rightskip0pt\parfillskip0pt plus1fil\vtop{\hsize102mm\noindent #1}}
\caption{List of distributions generated by sampling programs and available for download from the project web page.}
\label{tab:dists}
\begin{center}
\begin{tabular}{cl}\hline
$t$-distributions & \vtop{%
	\Box{%
		{\em available for all Coulomb-interference modes}:\hfil\break
		amplitude $F(t)$, phase $\arg F(t)$, rho parameter $\rho(t) \equiv \Re F(t)/\Im F(t)$,
		differential cross-section $\d\sigma/\d t$, cumulative cross-section and elastic slope $B(t) \equiv \d\log |F(t)|^2/\d t$
	}%
	\vskip2mm
	\Box{%
		{\em derived quantities}:\hfil\break
		$R(t)$, $Z(t)$ and $C(t)$, see Eqs. (\ref{e:R}) to (\ref{e:C})
	}%
} \cr\hline
$b$-distributions & \Box{profile function $A(b)$} \cr\hline
$s$-distributions & \Box{total cross-section $\sigma_{\rm tot}(s)$, rho parameter $\rho(s, t=0)$ and forward hadronic slope $B_0(s) \equiv \d\log |F(t)|^2/\d t |_{t = 0}$}\cr\hline
\end{tabular}
\end{center}
\end{table}

The are programs to sample many distributions of interest (see the list in Table \ref{tab:dists}): {\tt ElegentXDistributionSampler} where {\tt X} can be {\tt T} for distributions in four-momentum transfer squared, {\tt B} for distributions in impact parameter space or {\tt S} for distributions in cetre-of-mass energy $\sqrt s$. The users' manual for these programs can be found at \cite[Wiki section]{elegent}. There are also scripts ({\tt generate\_x\_distributions}) to generate the distributions at ISR, S$\rm p\bar p$S, RHIC, Tevatron and LHC energies. These distributions are available for download in the ROOT format at the project web page \cite[Distributions section]{elegent}.

A cumulative cross-section distribution (sampled by the program mentioned above) can be used for Monte-Carlo generation of elastic nucleon-nucleon scattering events. This functionality is provided by the {\tt Generator} class, which can easily be used in any other program. For an example application see the  {\tt ElegentTest} program.

The sampled distributions are plotted to compare hadronic models in each of the quantities. These plots are available in PDF format at the project web page \cite[Plots section]{elegent}.



\end{document}